\documentclass[letter,traditabstract]{aa} 
\usepackage{graphicx}

\usepackage{natbib}
\bibpunct{(}{)}{;}{a}{}{,} 
\usepackage{txfonts}

\begin{document}
   \title{Evidence of a thick disk rotation--metallicity correlation}

   \author{A. Spagna\inst{1},
          M.G. Lattanzi\inst{1},
          P. Re Fiorentin\inst{2},
          \and
          R.L. Smart\inst{1}
          }
   \institute{INAF--Osservatorio Astronomico di Torino,
             via Osservatorio 20, 10025 Pino Torinese, Italy\\
             \email{spagna@oato.inaf.it}
        \and
               University of Ljubljana, Faculty of Mathematics and Physics, Jadranska 19,
               SLO-1000 Ljubljana, Slovenia         
             }

   \date{Received October 23, 2009; accepted January 13, 2010}
   \authorrunning{A. Spagna et al.}

 
  \abstract
   {We analyze  a new kinematic survey that includes accurate proper motions derived from SDSS DR7
positions, combined with multi-epoch measurements from the GSC-II database. 
By means of the SDSS spectro-photometric data (effective temperature, surface gravity, metallicity, and radial velocities), 
we estimate photometric parallaxes for a sample of 27\,000 FGK (sub)dwarfs with [Fe/H]$<-0.5$, which we adopted as tracers of the 
seven-dimensional space distribution (kinematic phase distribution plus chemical abundance) of the thick disk and inner halo within a few kiloparsecs of the Sun.\\
We find evidence of a kinematics-metallicity correlation, $\partial \langle V_\phi \rangle/ \partial $[Fe/H]$\approx 40\div 50$ km~s$^{-1}$ dex$^{-1}$,  amongst thick disk stars located between one and three kiloparsecs from the plane and with abundance $-1<$[Fe/H]$<-0.5$, while no significant correlation is present for [Fe/H]$\ga -0.5$.  
In addition, we estimate a shallow vertical rotation velocity gradient, 
$\partial \langle V_\phi \rangle/ \partial \left| z\right| = -19 \pm 2$ km~s$^{-1}$ kpc$^{-1}$, 
for the thick disk between 1 kpc $<\left| z\right|< 3$ kpc, and a low prograde rotation, $37\pm 3$ km~s$^{-1}$ for the inner halo up to 4 kpc. \\
Finally, we briefly discuss the implications of these findings for the thick disk formation scenarios 
in the context of CDM hierarchical galaxy formation mechanisms and of secular
evolutionary processes in galactic disks.
 }


   \maketitle
%

\section{Introduction}

The existence of a thick disk in our Galaxy was revealed by  \cite{gilmore1983}, 
who analyzed starcounts towards the South Galactic Pole.  Thanks to the many studies carried out
since then, the main spatial, kinematic, and chemical features 
of this population are well established. 
Thick disks have been also observed in many disk galaxies \citep{yoachim2006}, and they represent
the frozen relics of the first phases of disk galaxy formation \citep{freeman2002}.
However, in spite of the many scenarios  proposed until now, the origin of this component is still unclear.

In the context of  CDM hierarchical galaxy formation models, it is possible that thick disks are formed by the heating of a pre-existing thin disk through a minor merger \citep[e.g.][]{villalobos2008}, by accretion of stars from disrupted satellites \citep{abadi2003}, or by the stars formed {\it in situ}
from gas-rich chaotic mergers at high redshift \citep{brook2005}. On the other hand, simulations suggest that 
thick disks could simply be produced through secular radial migration of stars induced by the spiral arms \citep{roskar2008, schonrich2009}.

In any event, most astronomers agree that our thick disk is formed of an old stellar population
 with an age of 8-12 Gyr \citep[e.g.][and references therein]{haywood2008}.
The bulk of the thick disk stars have metallicity in the range 
 $-1\la $[Fe/H]$\la -0.3$ ( [Fe/H]$\simeq -0.6$, on average) with enhanced [$\alpha$/Fe] \citep{bensby2005, reddy2006}, but note that tails with metal-poor stars down to [Fe/H]$\simeq -2$ \citep{chiba2000} and metal-rich stars up to [Fe/H]$\simeq 0$ \citep{bensby2007} have also been revealed.
Moreover, according to \citet{ivezic2008}, a mild vertical metallicity gradient shifts 
the mean metallicity to [Fe/H]$\simeq -0.8$ beyond $|z|\ga 3$ kpc. 

The spatial distribution is usually modeled with a symmetric exponential density distribution as a function of galactocentric coordinates $(R,z)$.
Its scale height spans a wide range of measurements, between $h_z=640$~pc and 1500~pc, while the local normalization varies beetween 13\% and 2\% in anticorrelation with $h_z$ 
\citep[see Fig. 3 of][]{arnadottir2008}.
The distribution above the galactic plane is supported by a vertical velocity dispersion, $\sigma_W\simeq 40$ km~s$^{-1}$, which is associated with an asymmetric drift of $\sim 50$ km~s$^{-1}$, relative to the local standard of rest.

Significant asymmetries have also been detected, such as the prominent Hercules thick disk cloud \citep{parker2003, juric2008}, which could correspond to a merger remnant or indicate a triaxial thick disk \citep{larsen2008}.

In this letter, we present new results regarding the vertical rotation gradient and, for the first time to our knowledge, evidence of a metallicity-rotation correlation in the thick disk stellar population.

\section{The SDSS -- GSC-II catalog}
 This study is based on a new kinematic catalog derived by assembling the astrometric 
parameters extracted from the database used for the construction of the Second Guide Star Catalog \citep[GSC-II; ][]{lasker2008} 
with spectro-photometric data from the Seventh Data Release of the Sloan Digital Sky Survey \citep[SDSS DR7; e.g.\ ][]{abazajian2009, yanny2009}.  
The SDSS--GSC-II catalog contains positions, proper motions, classification, and $ugriz$ photometry for 77 million sources down to $r\approx 20$,  over 9000 square-degrees. 

Proper motions are computed by combining multi-epoch positions from SDSS DR7 and the GSC-II database. Typically,  5-10 observations  are available for each source, spanning $\sim50$ years. 
 Total errors are in the range 2-3 mas~yr$^{-1}$ for $16<r<18.5$, comparable with those of the SDSS proper motions \citep{munn2004}, as confirmed by external comparisons against QSOs. 
The construction and properties of this catalog are described in detail by Smart et al. (2010, in preparation), while a concise description can be found in \citet{spagna2009}.

Radial velocities ($\sigma_{Vr}< 10$  km~s$^{-1}$) and astrophysical parameters ($\sigma_{\rm Teff}\simeq 150$ K, $\sigma_{\log g}\simeq 0.25$, $\sigma_{\rm [Fe/H]}\simeq 0.20$) are available for 151\,000 sources cross-matched with the SDSS spectroscopic catalog.
  From this list, we select sources with $4500$ K$<T_{\rm eff}< 7500$ K and $\log g> 3.5$, corresponding to FGK (sub)dwarfs, and apply the color thresholds from \citet{klement2009}
in order to remove turn-off stars.

 Spectro-photometric distances are computed by means of metallicity-dependent absolute magnitude relations, $M_r=f(g-i,{\rm [Fe/H]})$, from  \citet{ivezic2008}.
Here, the observed magnitudes are corrected for interstellar absorption via the extinction maps of 
\citet{schlegel1998}, while the spectroscopic [Fe/H] is used, instead of the photometric metallicity applied by \citet{ivezic2008}.

The mean distance of the sample is $\sim 2$ kpc, while most (92\%) of the sources are distributed between 0.5 kpc $< |z| < 3.5$ kpc and 6 kpc $<R<11$ kpc.  The typical accuracy of the $M_r$ calibration is  0.3 mag (random) and 0.1 mag (systematic), which corresponds to distance errors of $\Delta d/d=15$\% and 5\%, respectively. 
Finally, 3D velocities in the galactocentric reference frame, $(V_R, V_\phi, V_z)$, are derived by assuming $R_\odot=8$ kpc,  solar motion  $(U_\odot,V_\odot,W_\odot)$ from \citet{dehnen1998}, and local standard of rest velocity of 220 km~s$^{-1}$.

In order to produce an accurate sample, we select only stars with $(i)$ proper motion errors $<10$ mas~yr$^{-1}$ per component, 
$(ii)$ errors on the velocity components $<50$ km~s$^{-1}$, $(iii)$ total velocity $< 600$ km~s$^{-1}$, $(iv)$ distance $<5$ kpc,  and $(v)$ magnitude $13.5<g<20.5$.
Overall, the {\it kinematic} catalog contains 46\,000 stars; in the following sections a subsample of 27\,000 low metallicity dwarfs with $-3<$[Fe/H]$<-0.5$ will be used as {\it tracers} of the inner halo and thick disk and analyzed in details.

\section{Analysis and results}

\subsection{Vertical rotation gradient}
\label{sect:verticalGradient}
\begin{figure}[]
\resizebox{\hsize}{!}{\includegraphics[trim=0mm 1mm 0mm 1mm clip=true]{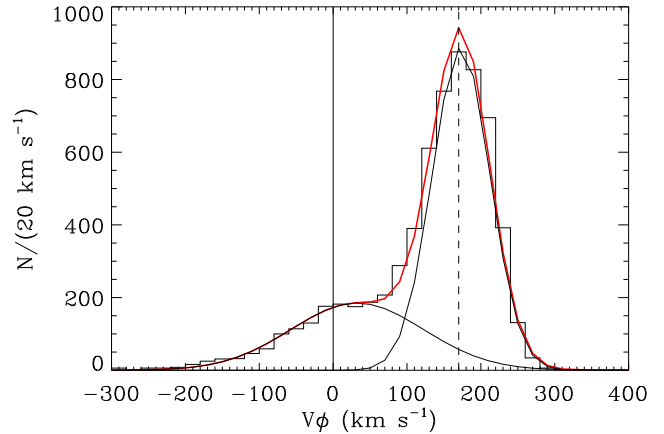}}
\caption{\footnotesize
Histogram of the velocity distribution, $V_\phi$, of the kinematic sample with [Fe/H]$<-0.5$, between $\left| z\right|=$1.0 kpc and 1.5 kpc. The thick solid line shows the best fit of a two Gaussian component model, representing the thick disk and halo populations (thin lines).}
\label{spagna_fig1}
\end{figure}

Figure \ref{spagna_fig1} shows the $V_\phi$ distribution of 6538 stars
with $1.0$ kpc $< \left|z\right|\le 1.5$ kpc and [Fe/H]$<-0.5$. In this sample, the contamination of thin disk stars is expected to be negligible\footnote
{Assuming a standard model with a thin disk and a thick disk having scale-heights of 300 pc and 900 pc, respectively, and a thick disk normalization of 10\% at $z$=0 pc,  about half of the stars belongs to the thin disk for $1.0$ kpc $<\left|z\right|< 1.5$ kpc, but only a few percent of them with [Fe/H]$<-0.5$ \citep[cfr. e.g.][]{aumer2009}. Also, we estimate the contamination of metal poor thin disk stars does not exceed 10\%, even if we adopt a thick disk with a shorter $h_z=580$ pc and a local normalization of 13\% \citep{chen2001}. }, so that 
we fit the distribution with only two gaussian populations, corresponding to the thick disk and halo.
The least-squares solution of the two-component model is good, although the counts at $V_\phi\approx$ 220 km~s$^{-1}$  are slightly underestimated ($\sim-16$\%) and the velocity peak is overestimated of about 7\%; 
this explains a non-optimal $\chi^2_\nu=3.18$.
 (If we force a third gaussian component corresponding to the thin disk, the formal goodness of fit improves significantly, $\chi^2_\nu=1.37$, but the solution becomes ill-conditioned with an inaccurate thin disk normalization of $(19\pm 6)$\%. )

\begin{table}
\caption{Parameters of a two-component Gaussian best fit (thick disk and halo) for six height intervals. }
\label{table:1}      
\begin{center}
\begin{tabular}{crllllcc}

\hline\hline

     &  & \multicolumn{2}{c}{\sc Thick Disk} & \multicolumn{2}{c}{\sc Halo} &  &  \\
 $\langle|z|\rangle$ & N &  $\langle V_\phi\rangle$  &  $\sigma_{V\phi}$ & $\langle V_\phi\rangle$  &  $\sigma_{V\phi}$ & $\frac{\rho_{\rm TD}}{\rho_{\rm tot}}$ & $\chi^2_\nu$\\
 (kpc) &  & \multicolumn{2}{c}{(km s$^{-1}$)} & \multicolumn{2}{c}{(km s$^{-1}$)} & (\%) & \\
\hline
 0.76 & 7022 & 186$\pm$1& 34$\pm$1 & 46$\pm$5 & 92$\pm$3 & 74$\pm$2 & 5.18\\  %

 1.24 & 6538 & 173$\pm$1& 39$\pm$1 & 32$\pm$4 & 90$\pm$3 & 68$\pm$2 & 3.18\\  
 1.73 & 4753 & 163$\pm$1& 44$\pm$2 & 29$\pm$11 & 96$\pm$5 & 60$\pm$4 & 1.42\\ 

 2.23 & 3044 & 155$\pm$2& 47$\pm$3 & 36$\pm$10 & 90$\pm$4 & 48$\pm$5 & 1.83\\ %

 2.73 & 1637 & 144$\pm$4& 42$\pm$5 & 49$\pm$6 & 97$\pm$3 & 29$\pm$6 & 1.14\\

 3.36 & 988 & 166$\pm$11 & 41$\pm$11& 44$\pm$8 &  90$\pm$4 & 13$\pm$7 & 0.80\\
\hline
\end{tabular}
\end{center}
\end{table}

\begin{figure*}[]
\begin{center}
\resizebox{0.90\hsize}{!}{\includegraphics[trim=0mm 0mm 0mm 0mm clip=true]{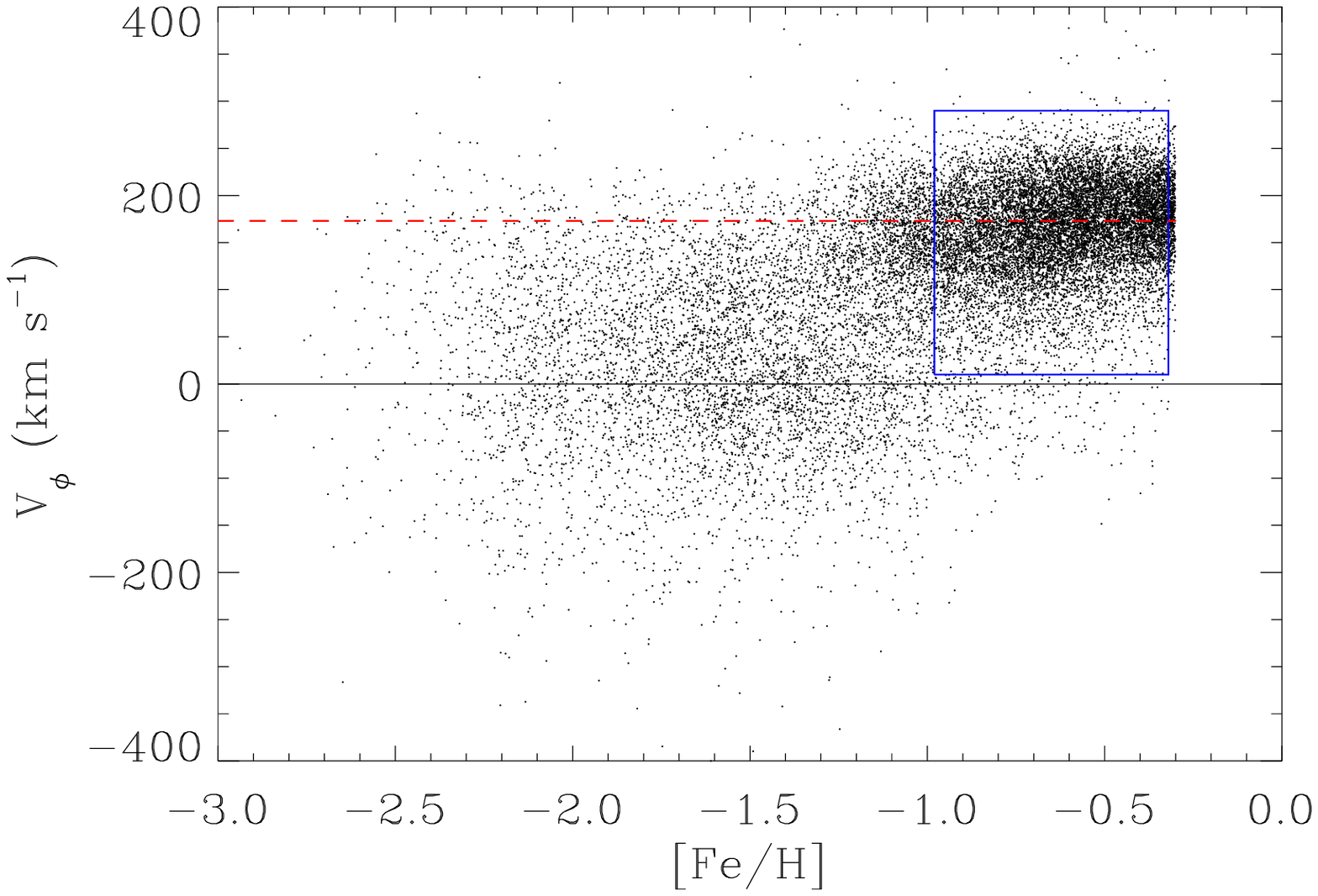}}
\caption{Velocity-metallicity distribution  of  20\,251 stars 
with $\left|z\right|=$1.0-3.0 kpc and [Fe/H]$<-0.3$. 
The dashed line indicates the thick disk rotation, $V_\phi=173$ km~s$^{-1}$ at $\langle \left|z\right|\rangle = 1.24$ kpc (Table \ref{table:1}). 
The box defines the region, shown in Fig. \ref{spagna_fig3}, in which the thick disk population dominates.}
\label{spagna_fig2}
\resizebox{0.80\hsize}{!}{\includegraphics[trim=0mm 0mm 0mm 0mm clip=true]{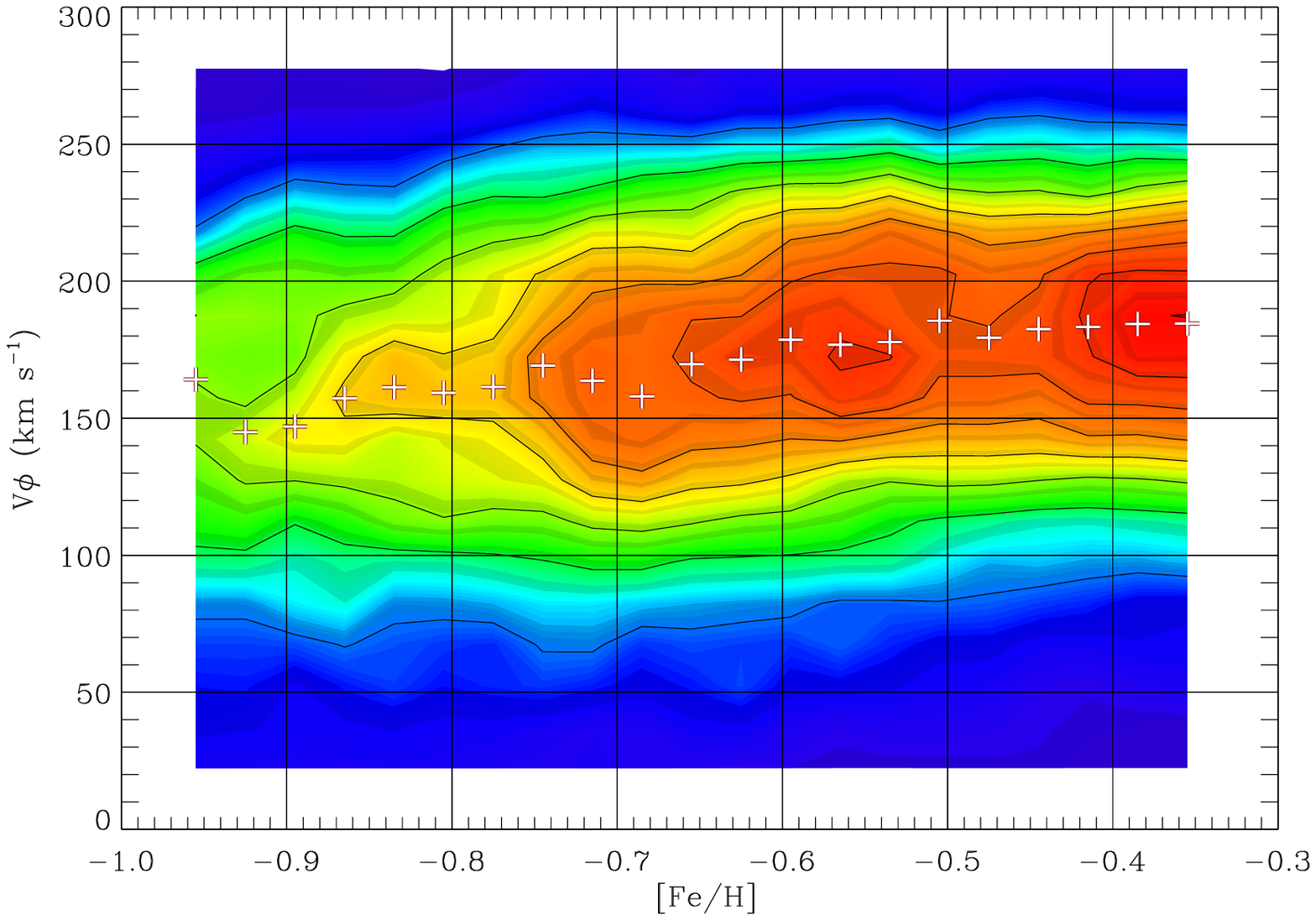}}
\caption{
\footnotesize
Iso-density contours of 13\,108 relatively metal-poor stars with $-1.0<$[Fe/H]$<-0.3$ and $\left|z\right|=$1.0-3.0 kpc. White crosses mark the ridge line of the maximum likelihood $V_\phi$ vs. [Fe/H].}
\label{spagna_fig3}
\end{center}
\end{figure*}

\begin{figure*}[]
\resizebox{0.90\hsize}{!}{\includegraphics[clip=false]{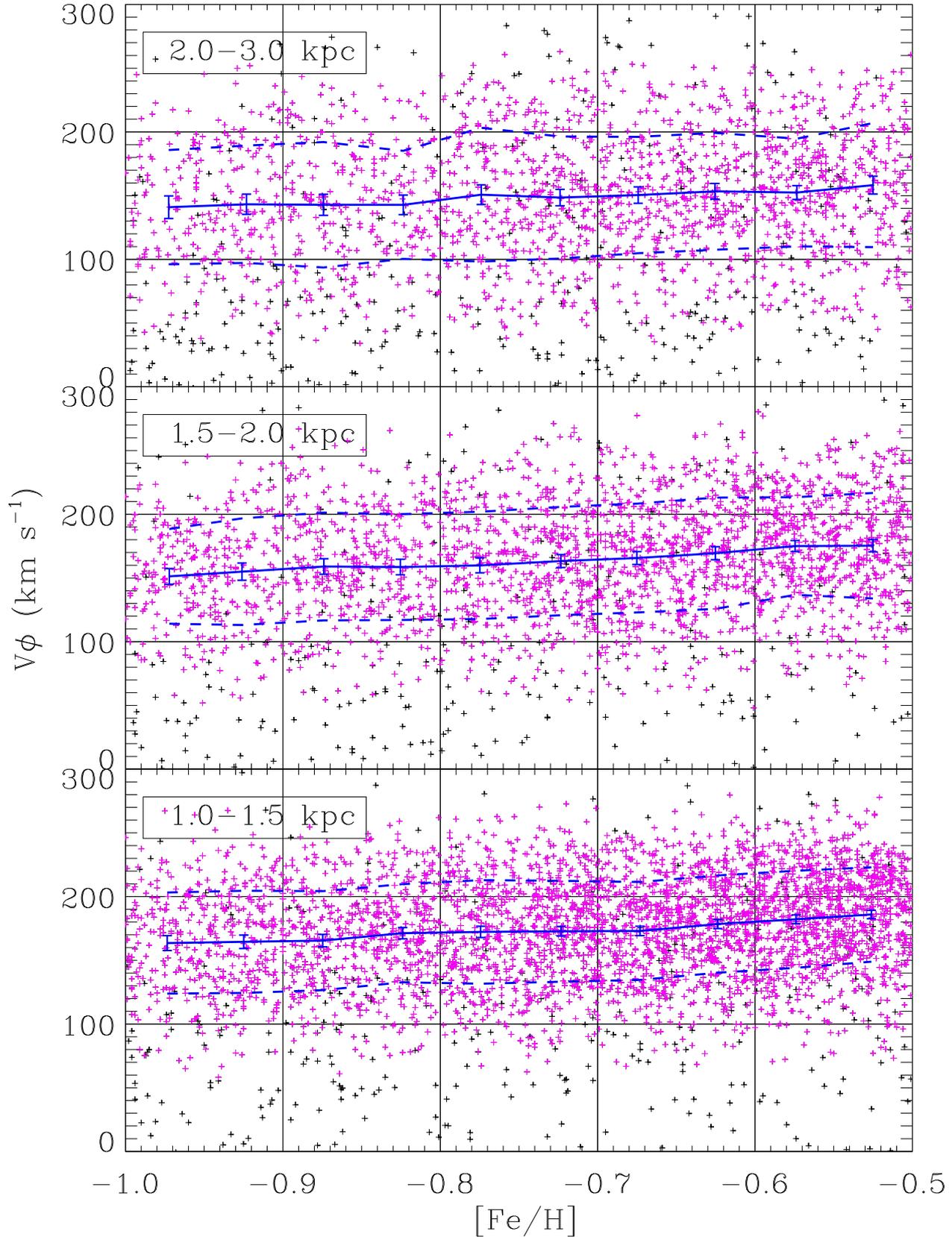}}
\caption{ $V_\phi$ vs.\ [Fe/H] distribution of stars with $-1.0<$[Fe/H]$<-0.5$.
 Black crosses mark the rejected stars beyond 3$\sigma$ of the velocity ellipsoid of the thick disk to minimize the contamination from halo stars (see Sect. \ref{sect:correlation}). The solid line connects the mean velocities $\langle V_\phi\rangle$, which are plotted with 2$\sigma$ error bars. The dashed lines indicate the $\pm 1\sigma$ spread of the velocity distribution.
 Top, middle, and bottom panels refer to $\left| z\right|=$1.0--1.5 kpc, 1.5--2.0 kpc, and 2.0--3.0 kpc, respectively.}
\label{spagna_fig4}
\end{figure*}

The same procedure is repeated for six height bins: $\Delta\left| z\right| =0.5$--1.0 kpc, 1.0--1.5 kpc, 1.5--2.0 kpc, 2.0--2.5 kpc, 2.5--3.0 kpc, and 3.0--4.0 kpc. The results are reported in Table \ref{table:1}, which lists mean height, number of stars, mean rotation velocities and dispersions, fraction of thick disk stars, and reduced $\chi^2_\nu$.
The halo parameters appear quite stable:  on average, $V_{\phi}\simeq 37 \pm 3$ km~s$^{-1}$ ($1<\left| z\right|\le 4$ kpc),  which indicates a slow prograde rotation of the inner halo, in agreement with some authors \citep{chiba2000,kepley2007} but different from others that favor a non-rotating inner halo \citep{vallenari2006, smith2009, bond2009}. 
The halo velocity dispersion also appears rather constant up to $\left|z\right|\simeq 4$ kpc, with a mean value 
of $\sigma_{V\phi}=93\pm 2$ km~s$^{-1}$ (uncorrected for the velocity errors).
Conversely, the thick disk shows a monotonic decreasing of the rotation velocity from 
$V_\phi=186$ km~s$^{-1}$ to 146 km~s$^{-1}$, for height from 0.5 kpc to 3 kpc. 
In the highest bin (3 kpc $\le \left|z\right|< 4$ kpc), $V_\phi$ increases to 166$\pm 11$ km~s$^{-1}$, but we think this is a spurious effect of both the larger velocity errors and the small fraction, $(13\pm 7)$\%, of thick disk stars that are strongly entangled with the halo population. 
Similarly, in the same $z$-range, the velocity dispersion increases from $\sigma_{V\phi}=34$  km~s$^{-1}$  to $\sim$45  km~s$^{-1}$, in part because of the tangential velocity errors that scale with distance.

We exclude the highest bin and also the lowest, as it is probably contaminated by thin disk stars which are difficult to deconvolve from the thick disk population. 
Thus, we estimate the gradient,
\begin{equation}
{\partial \langle V_{\phi}\rangle}/{\partial \left|z\right|} = -19 \pm 2 \hbox{\rm { } km~s$^{-1}$ kpc$^{-1}$}
\end{equation}
and the extrapolated intercept,  $V_{\phi}(z=0)=196\pm 3$ km~s$^{-1}$.
Our result is significantly smaller than the value, $-30\pm 3$ km~s$^{-1}$~kpc$^{-1}$,  measured by 
 \citet{chiba2000}, who analyzed stars with abundance in the range, $-0.8\le$[Fe/H]$\le -0.6$, where the thick disk dominates. A similar trend was estimated by \citet{girard2006}, \citet{carollo2009}, and by \citet{bond2009}, although they adopted a nonlinear function.  

Instead, a shallower slope was found by \citet{majewski1992}, who derived a gradient of $-21\pm 1$ km~s$^{-1}$~kpc$^{-1}$ for $\left|z\right|<5$ kpc, after separating the halo population from that of the thick disk.
A low kinematical gradient was also found by \citet{spagna1996} and, more recently, by  \citet{allende-prieto2006}, who estimated $-10$ km~s$^{-1}$~kpc$^{-1}$ and $-16$ km~s$^{-1}$~kpc$^{-1}$, respectively. 
 The difference between these results can be explained, at least in part, by thin disk and halo stars contamination, which tends to produce steeper velocity gradients.

\begin{table}
\caption{Kinematics-metallicity correlation of thick disk stars with $-1.0<$[Fe/H]$<-0.5$ for three height intervals.}
\label{table:2}      
\begin{center}
\begin{tabular}{ccccc}
\hline\hline
   $\langle \left|z\right| \rangle$ & N$_{\rm tot}$ & N$_{\rm used}$ &  $\partial \langle V_\phi\rangle / \partial$[Fe/H] & $\rho_s$ \\
 (kpc) &  &  & (km s$^{-1}$ dex$^{-1}$) &  ($\times 10^{-2}$)\\
      &      &   $3\sigma$ {  } { } $2\sigma$ &  $3\sigma$ { } { } { }  $2\sigma$ &  $3\sigma$ { } { } { }  $2\sigma$\\
\hline
 1.23 & 3994 & 3672 { } 2915 & 50$\pm$5 { } 39$\pm$5 & 17$\pm$2 { } 15$\pm$2 \\   
 1.73 & 2641 & 2348 { } 1715 & 54$\pm$6 { } 35$\pm$5 &  18$\pm$2 { } 16$\pm$2 \\ 
 2.37 & 2194 & 1768 { } 1131 & 35$\pm$8 { } 33$\pm$5 &  10$\pm$2 { } 14$\pm$3\\  
\hline
\end{tabular}
\end{center}
\end{table}
\begin{figure}[]
\resizebox{\hsize}{!}{\includegraphics[trim=0mm 3mm 0mm 2mm]{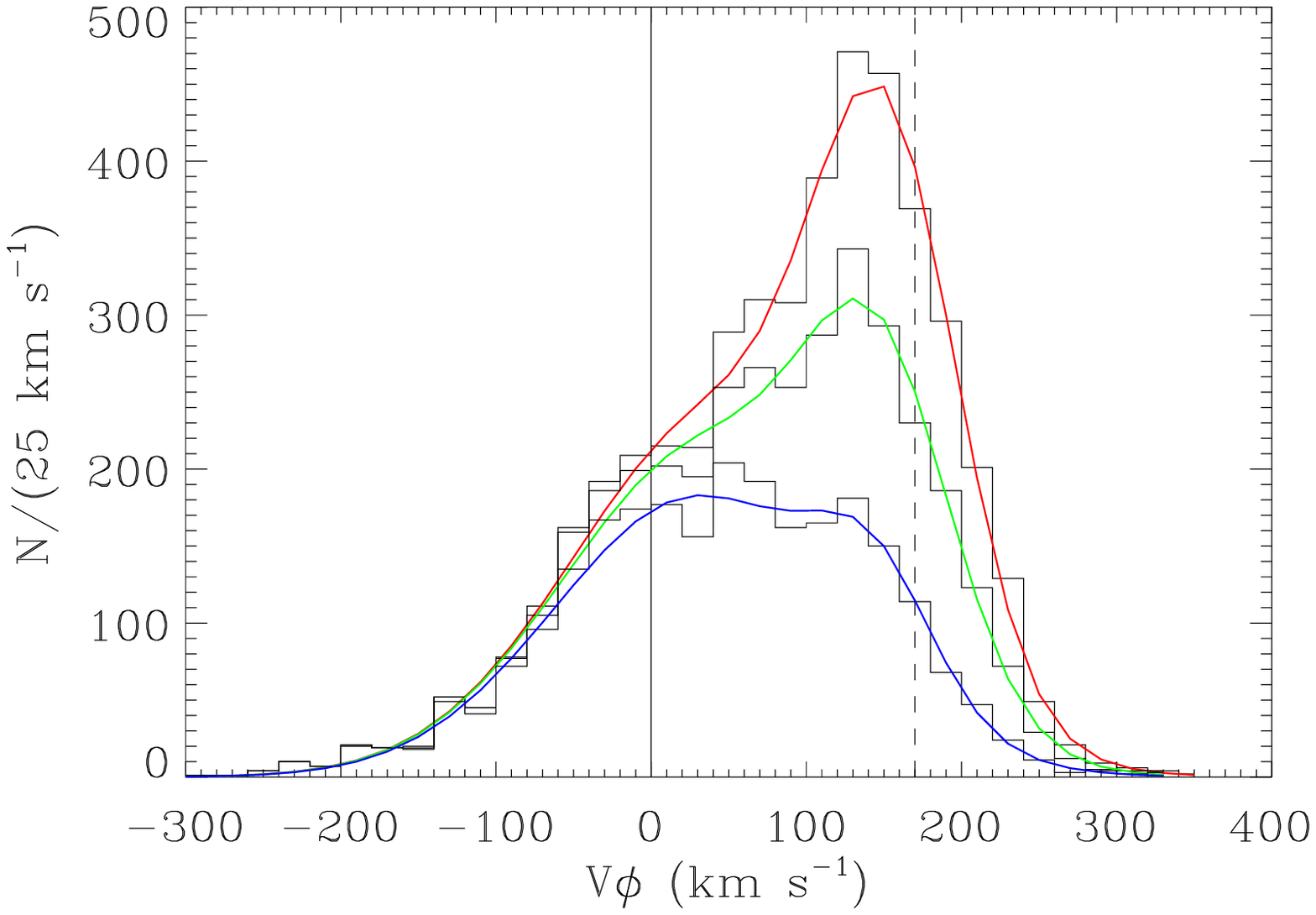}}
\resizebox{\hsize}{!}{\includegraphics[trim=0mm 3mm 0mm 2mm]{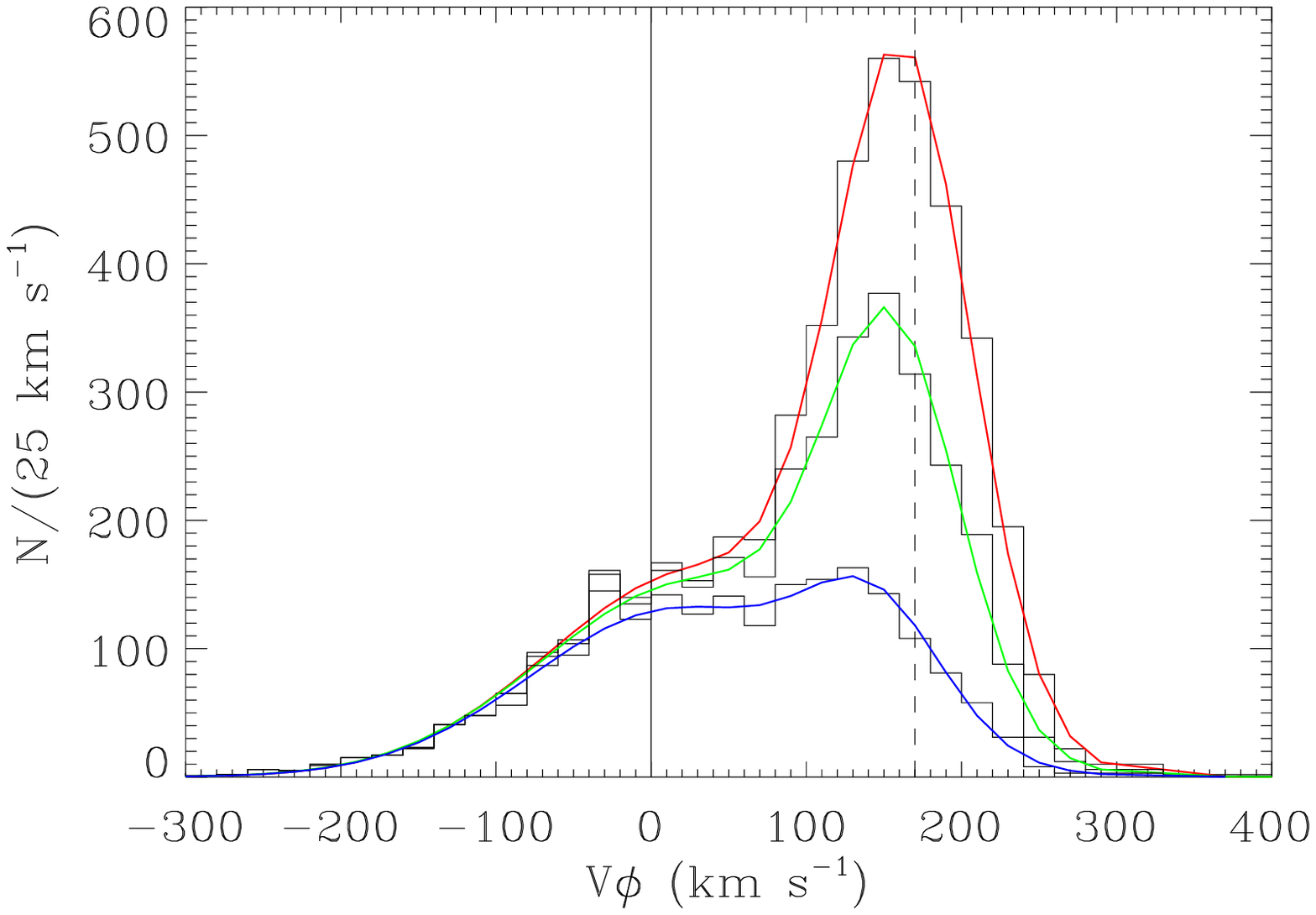}}
\resizebox{\hsize}{!}{\includegraphics[trim=0mm 3mm 0mm 2mm]{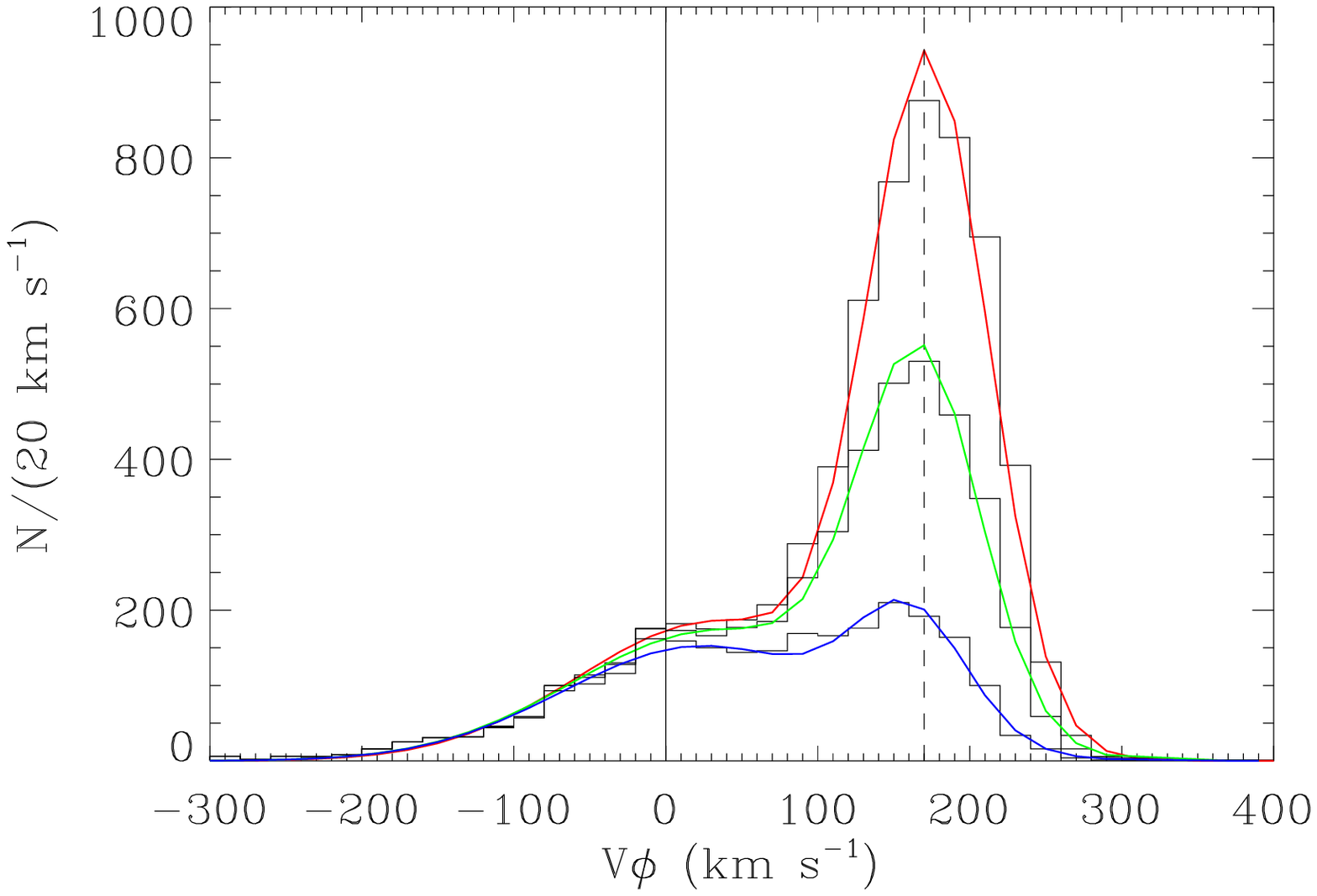}}
\caption{$V_\phi$ distribution for [Fe/H]$<-0.5$,  $-0.7$, and $-1.0$, in descending order. The solid lines show the best fit of two Gaussian-component models (thick disk and halo). Bottom, middle, and top panels refer to $\left| z\right|=$1.0--1.5 kpc, 1.5--2.0 kpc, and 2.0--3.0 kpc, respectively. The solid and dashed lines mark the values $V_\phi=0$ km~s$^{-1}$ and 170 km~s$^{-1}$. }
 \label{spagna_fig5}
\end{figure}
\subsection{Rotation -- metallicity correlation}
\label{sect:correlation}
The {\it disk} and {\it halo} populations are apparent in the $V_\phi$ vs.\ [Fe/H] distribution (see Fig.\ \ref{spagna_fig2}).
In particular, the region $-1.0<$[Fe/H]$<-0.5$ and 0 km~s$^{-1}<V_\phi<300$ km~s$^{-1}$ does contain the bulk of the regular thick disk stars, besides a small number of stars belonging to the metal-poor tail of the thin disk and to the high-metallicity tail of the inner halo. Actually, a significant fraction of thin disk stars are expected for [Fe/H]$>-0.5$, while towards lower abundances, [Fe/H]$<-1$, the thick disk metal weak tail and the newly discovered flattened inner halo \citep{morrison2009} are also present.

Figure \ref{spagna_fig3} shows the iso-density contours of the velocity-metallicity distribution of stars with  $\left|z\right|=1.0$-3.0 kpc and  $-1.0<$[Fe/H]$<-0.3$. 
As in \citet{ivezic2008} and \citet{bond2009}, no correlation appears in the transition region between the thin and thick disks ( [Fe/H]$\ga -0.5$).  
Instead, we notice a shallow but clear slope for [Fe/H]$\la -0.5$, undetected by previous studies, which indicates that the metal-rich stars tend to rotate faster than the metal-poor ones. 
In particular, the top-density ridge increases from $V_\phi\simeq 150$ km~s$^{-1}$ at [Fe/H]$\approx -1$ to 
$V_\phi\simeq 170$ km~s$^{-1}$ at [Fe/H]$\approx -0.4$.
Inspection of Fig. \ref{spagna_fig3} also proves a bimodal distribution with a secondary maximum located at [Fe/H]$\approx -0.55$, close to the value of the mean metallicity of the thick disk, and the peak at [Fe/H]$\simeq -0.38$ due to thin disk stars.

\begin{table*}
\caption{Fitted parameters, as in Table \ref{table:1}, for different metallicity intervals, $-3.0<$ [Fe/H] $\le$ [Fe/H]$_\mathrm{max}$, where $-1.0\le$ [Fe/H]$_\mathrm{max} \le -0.5$.  }
\label{table:3}      
\begin{center}
\begin{tabular}{crllllcc}
\hline\hline
     &  & \multicolumn{2}{c}{\sc Thick Disk} & \multicolumn{2}{c}{\sc Halo} &  &  \\
 $\left[\mathrm{Fe}/{\mathrm H}\right]_\mathrm{max}$   & N &  $\langle V_\phi\rangle$  &  $\sigma_{V\phi}$ & $\langle V_\phi\rangle$  &  $\sigma_{V\phi}$ & $\frac{\rho_{\rm TD}}{\rho_{\rm tot}}$ & $\chi^2_\nu$\\
 (dex) &  & \multicolumn{2}{c}{(km s$^{-1}$)} & \multicolumn{2}{c}{(km s$^{-1}$)} & (\%) & \\
\hline
            \multicolumn{8}{c}{$1.0$ kpc $<\left|z\right| \le 1.5$ kpc}\\
\hline
 $-0.5$ & 6537 & 173 $\pm$ 1& 39 $\pm$ 1 & 33 $\pm$ 4 & 90 $\pm$ 3 & 68 $\pm$ 2 & 3.19\\
 $-0.6$ & 5470 & 170 $\pm$ 1& 39 $\pm$ 1 & 35 $\pm$ 7 & 94 $\pm$ 4 & 61 $\pm$ 3 & 2.13\\
 $-0.7$ & 4511 & 167 $\pm$ 1& 39 $\pm$ 2 & 33 $\pm$ 7 & 94 $\pm$ 4 & 55 $\pm$ 3 & 1.80\\
 $-0.8$ & 3675 & 165 $\pm$ 1& 38 $\pm$ 3 & 31 $\pm$ 6 & 93 $\pm$ 3 & 46 $\pm$ 3 & 1.57\\
 $-0.9$ & 3036 & 162 $\pm$ 2& 38 $\pm$ 2 & 29 $\pm$ 6 & 93 $\pm$ 3 & 38 $\pm$ 4 & 1.51\\
 $-1.0$ & 2543 & 162 $\pm$ 2& 37 $\pm$ 2 & 24 $\pm$ 5 & 92 $\pm$ 3 & 30 $\pm$ 4 & 1.59\\
\hline
            \multicolumn{8}{c}{$1.5$ kpc $<\left|z\right| \le 2.0$ kpc}\\
\hline
 $-0.5$ & 4753 & 163 $\pm$ 1& 44 $\pm$ 2 & 29 $\pm$ 10 & 96 $\pm$ 5 & 60 $\pm$ 4 & 1.42\\
 $-0.6$ & 4113 & 157 $\pm$ 2& 45 $\pm$ 2 & 23 $\pm$ 10 & 94 $\pm$ 5 & 56 $\pm$ 4 & 1.49\\
 $-0.7$ & 3480 & 154 $\pm$ 2& 45 $\pm$ 2 & 23 $\pm$ 11 & 95 $\pm$ 5 & 49 $\pm$ 5 & 1.28\\
 $-0.8$ & 2936 & 152 $\pm$ 2& 43 $\pm$ 3 & 22 $\pm$ 9  & 95 $\pm$ 4 & 41 $\pm$ 5 & 1.30\\
 $-0.9$ & 2488 & 149 $\pm$ 3& 42 $\pm$ 3 & 19 $\pm$ 8  & 94 $\pm$ 4 & 33 $\pm$ 5 & 1.20\\
 $-1.0$ & 2112 & 145 $\pm$ 4& 44 $\pm$ 5 & 18 $\pm$ 9  & 94 $\pm$ 4 & 25 $\pm$ 6 & 1.18\\
\hline
            \multicolumn{8}{c}{$2.0$ kpc $<\left|z\right| \le 3.0$ kpc}\\
\hline
 $-0.5$ & 4680 & 152 $\pm$ 2& 45 $\pm$ 2 & 43 $\pm$ 6 & 94 $\pm$ 2 & 40 $\pm$ 4 & 1.72\\
 $-0.6$ & 4176 & 150 $\pm$ 3& 46 $\pm$ 3 & 42 $\pm$ 6 & 94 $\pm$ 2 & 34 $\pm$ 4 & 1.56\\
 $-0.7$ & 3671 & 147 $\pm$ 3& 45 $\pm$ 3 & 37 $\pm$ 6 & 93 $\pm$ 2 & 30 $\pm$ 4 & 1.41\\
 $-0.8$ & 3183 & 144 $\pm$ 4& 43 $\pm$ 4 & 33 $\pm$ 5 & 92 $\pm$ 2 & 25 $\pm$ 4 & 1.31\\
 $-0.9$ & 2830 & 143 $\pm$ 4& 40 $\pm$ 5 & 32 $\pm$ 5 & 92 $\pm$ 2 & 19 $\pm$ 4 & 1.20\\
 $-1.0$ & 2486 & 141 $\pm$ 5& 38 $\pm$ 6 & 30 $\pm$ 4 & 93 $\pm$ 2 & 14 $\pm$ 4 & 1.20\\
\hline
\end{tabular}
\end{center}
\end{table*}

To quantify the correlation, we first select the stars within $\Delta$[Fe/H]=0.05 bins 
in the range $-1.0<$[Fe/H]$< -0.5$ and located at the different height intervals: $\Delta\left| z\right|= 1.0$--1.5 kpc, 1.5--2.0 kpc, and 2.0--3.0 kpc.  
Then, the stars with velocities $(V_R, V_\phi, V_z)$ outside 3$\sigma$ from the thick disk velocity ellipsoid, corresponding to a confidence level of 97.1\%, were rejected to minimize the contamination from the halo stars. 
   We adopted $\langle V_\phi\rangle$ as a function of $z$ derived from Table \ref{table:1} and assumed constant dispersions: $\sigma_{V\phi}\equiv\sigma_{Vz}= 40$ km~s$^{-1}$ and $\sigma_{V_R}\equiv 1.5\, \sigma_{V\phi}= 60$ km~s$^{-1}$. 
  
Finally, mean velocities were computed for the {\it bona fide} thick disk stars and the slope, $\partial \langle V_\phi\rangle / \partial$[Fe/H], is estimated by means of a linear fit for the height intervals $\Delta\left| z\right| =$1.0--1.5 kpc, 1.5--2.0 kpc, and 2.0--3.0 kpc. For each bin, mean height, total number of stars,  number of stars used (after 3$\sigma$ and 2$\sigma$ rejection), slope, and Spearman's rank correlation coefficient are listed in Table \ref{table:2}, while the observed distributions are shown in Fig. \ref{spagna_fig4}.
Overall, a kinematic-metallicity correlation of about 50 km~s$^{-1}$~dex$^{-1}$  is detected up to $ \left| z\right|\simeq 2$ kpc, while a shallower slope ($\sim35$ km~s$^{-1}$~dex$^{-1}$)  is present between $2 < \left| z\right|\le 3$ kpc.  
It is possible that these values are affected by a residual contamination of halo stars, whose presence can be inferred by the number of rejected high velocity stars shown in Table \ref{table:2} being greater than the 3\% expected in the case of a pure Gaussian distribution.  Nevertheless, even if we apply a conservative 2$\sigma$ selection (73.8\% confidence level), we still find a correlation at the level of 30$\div$40 km~s$^{-1}$~dex$^{-1}$, as reported in the last column of Table \ref{table:2}.

This conclusion is consistent with the systematic slowing down of the thick disk rotation, which results from fitting a {\it two} Gaussian-component model, representing the thick disk and halo populations, as more metal-poor thresholds are applied: [Fe/H]$_{\mathrm{max}}<-0.5, <-0.6,$ ... $<-1.0$ (see Table \ref{table:3}). This effect is depicted in Fig. \ref{spagna_fig5}, which shows how the thick disk component both decreases {\it and} shifts towards lower $V_\phi$ values, when different subsamples of metal poor stars are selected.

\begin{figure}[]
\resizebox{\hsize}{!}{\includegraphics[trim=0mm 3mm 0mm 3mm]{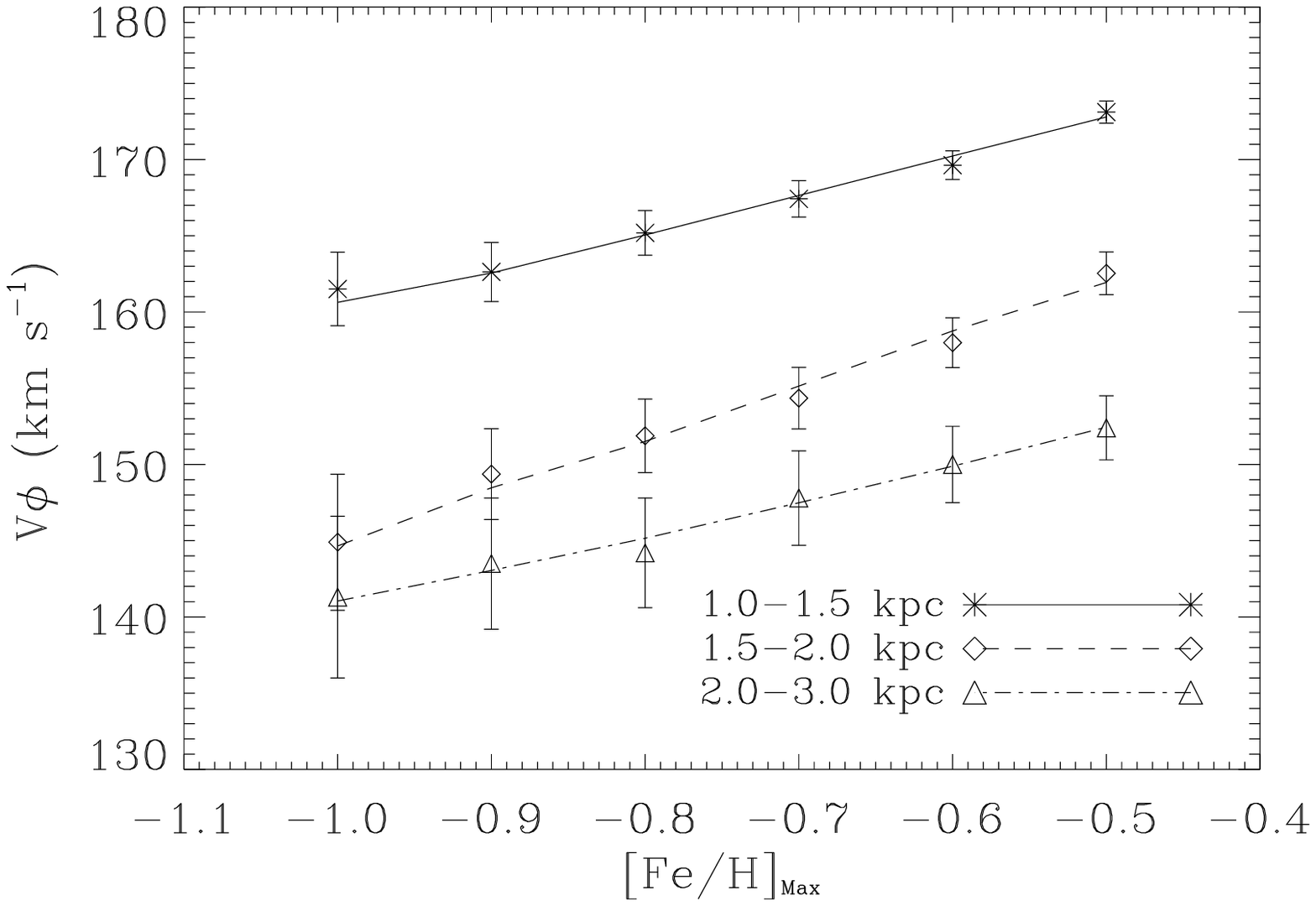}}
\caption{
\footnotesize
Mean $V_\phi$ vs. [Fe/H]$_{\rm max}$ for different abundance ranges, 
$-3< $[Fe/H]$\le $[Fe/H]$_{\rm max}$, with $-1.0\le $[Fe/H]$_{\rm max} \le -0.5$. The three lines show the $\langle V_\phi\rangle$ computed from Eq. \ref{eq:2} with fitted slopes,  $\partial \langle V_\phi\rangle / \partial$[Fe/H]= 
 $( 39 \pm 5,  56 \pm 9, 38\pm 13)$ km s$^{-1}$ dex$^{-1}$, and zero-points,  
$V_\phi(\mathrm{[Fe/H]_0})=(178\pm 1,  170\pm 2, 158\pm 3)$ km s$^{-1}$ for $\left|z\right|=$(1.0-1.5, 1.5-2.0, 2.0-3.0) kpc, respectively.}
\label{spagna_fig6}
\end{figure}

In addition, we estimate the rotation-metallicity correlation by fitting the thick disk $\langle V_\phi\rangle$ values from Table \ref{table:3} through the following integral linear model:
\begin{equation}
 \langle V_\phi \rangle =  V_\phi(\mathrm{[Fe/H]}_0) + a\cdot \left( \langle \mathrm{[Fe/H]}\rangle - \mathrm{[Fe/H]}_0 \right),
 \label{eq:2}
\end{equation}
where $\langle \mathrm{[Fe/H]}\rangle$ is the average for the stars with $-3<$ [Fe/H]$\le $[Fe/H]$_{\rm max}$, $a=\partial \langle V_\phi\rangle/\partial\mathrm{[Fe/H]}$, and $V_\phi$([Fe/H]$_0$) is the mean velocity of the reference metallicity, which we set to [Fe/H]$_0=-0.6$.
In Figure \ref{spagna_fig6}, the lines connect the values from Eq. \ref{eq:2} at the different [Fe/H]$_{\mathrm{max}}$ thresholds.   These results confirm both a vertical gradient consistent with the value derived in Sect.\ \ref{sect:verticalGradient} and a rotation-metallicity correlation in the range of 40$\div$50 km~s$^{-1}$~dex$^{-1}$ for the thick disk.

We also considered the hypothesis that a false trend  $V_\phi$ vs.\ [Fe/H] might derive from the tangential velocity estimated through the metallicity-dependent photometric parallaxes. Actually, the correlation would still be  significant even if the $M_r$-calibration were subjected to a systematic error up to 0.4 mag per dex.
Moreover, no kinematics-metallicity correlation is expected to arise because of the 
color-selection criteria of the SDSS spectroscopic targets, which although they produce a bias towards metal poor stars, cannot affect the conditional $V_\phi$ probability distribution at a given metallicity, Pr($V_\phi \left|\right.$[Fe/H]), and no further kinematical selection is applied.\\ 
 Thus, we conclude that the observed correlation is an intrinsic signature of our sample.
%
%
%
\section{Discussion and conclusions}
The existence of a vertical velocity gradient and a rotation-metallicity correlation sets important constraints on the origin of the thick disk. 
The estimated gradient of $-19\pm 2$ km~s$^{-1}$ kpc$^{-1}$ is consistent with Nbody simulations of disks thickened  by a single minor merger with a low/intermediate orbital inclination \citep[e.g.][]{villalobos2008}, 
as well as by the interaction with numerous dark subhalos, as discussed by \citet{hayashi2006} and \citet{kazantzidis2008}, 
whose simulations show kinematic gradients of $-(10\div 30)$ km~s$^{-1}$~kpc$^{-1}$ and of $-20$ km~s$^{-1}$~kpc$^{-1}$, respectively, for 1 kpc $<\left|z\right|\le 3$ kpc.
A vertical rotation gradient of about  $-20$ km~s$^{-1}$~kpc$^{-1}$ can also be inferred from Fig.\ 5 of \citet{abadi2003}, 
who investigated thick disks formed by accretion of both the stars of a pre-existing thin disk and the debris from disrupted satellites.
Unfortunately, we have not found any explicit kinematic prediction in the scenario of the chaotic gas-rich mergers described by \citet{brook2005}, 
although \citet{hayashi2006} state that a velocity shear {``may have difficulties in this regard''}.
Finally, to the best of our knowledge,  explicit predictions of kinematics-metallicity correlations 
are missing in the current CDM scenarios of satellite accretion or minor mergers. Hopefully, our results will motivate theoreticians to investigate this issue in their future models.

In the context of  models based on disk secular processes of stellar migration driven by interactions with spiral arms, a vertical gradient of $\sim -15$ km~s$^{-1}$~kpc$^{-1}$ is reported by \citet{loebman2008}, who, conversely, 
did not detected any $V_\phi$ vs.\ [Fe/H] correlation. The simulations carried out by \citet{schonrich2009} indicate a mild trend ($\sim10$ km~s$^{-1}$~dex$^{-1}$) at $z\approx 0$ kpc, which decreases with height and disappears for $\left| z\right|\ga 1$ kpc. Possibly, by adopting appropriate parameters, their inside-out disk formation model could reproduce the observed downtrend (Sch\"{o}nrich, 2009, private communication). 
Thus, more attention should be devoted to this scenario as a possible theoretical framework to explain the rotation--metallicity relation in the thick disk of the Milky Way.

\begin{acknowledgements}
We are grateful to the anonymous referee for all the valuable comments.
A.S. thanks Beatrice Bucciarelli and Ralph Sch\"{o}nrich for helpful discussions.
We acknowledge B. McLean and the GSC-II team for supporting the data mining of the GSC-II database. 
 The authors acknowledge the financial support of INAF through the PRIN 2007 grant n. CRA 1.06.10.04 ``The local route to galaxy
formation''. 
Support through the Marie Curie Research Training Network ELSA under contract MRTN-CT-2006-033481 to P.R.F. is also thankfully acknowledged.
Funding for the SDSS and SDSS-II has been provided by the Alfred P. Sloan Foundation, the Participating Institutions, the National Science Foundation, the U.S. Department of Energy, the National Aeronautics and Space Administration, the Japanese Monbukagakusho, the Max Planck Society, and the Higher Education Funding Council for England. The Guide Star Catalogue~II is a joint project of the Space Telescope Science Institute and the Osservatorio Astronomico di
Torino. 
\end{acknowledgements}

\bibliographystyle{aa}

\end{document}